\documentclass[twocolumn,superscriptaddress,showpacs,preprintnumbers,amsmath,amssymb,nofootinbib]{revtex4-1}

\usepackage{latexsym}% Basic package(s)
\usepackage{graphicx}% Include figure files
\usepackage{dcolumn}% Align table columns on decimal point
\usepackage{slashbox}% LaTeX table cell with a diagonal line
\newcommand{\bra}[1]{\langle #1 |}
\newcommand{\ket}[1]{| #1 \rangle}
\newcommand{\braket}[2]{\langle #1 | #2 \rangle}
\newcommand{\im}{\dot{\iota}\,}
\newcommand{\vecI}{\mathrm{I}}
\newcommand{\ketp}[1]{\ket{#1_+}}
\newcommand{\ketm}[1]{\ket{#1_-}}

\newcommand{\ketv}[1]{\ket{\mathbf{v}_{#1}}}
\newcommand{\brav}[1]{\bra{\mathbf{v}_{#1}}}
\newcommand{\braketv}[2]{\langle \mathbf{v}_{#1} | \mathbf{v}_{#2} \rangle}
\begin{document}

\title{Dynamical quantum repeater using cavity-QED and optical coherent states}

\author{Denis Gon\c{t}a}
\email{denis.gonta@mpl.mpg.de}
\affiliation{Institute of Optics, Information and Photonics,
             Friedrich-Alexander-University Erlangen-Nuremberg,
             Staudtstrasse 7, 91058 Erlangen, Germany}

\author{Peter van Loock}
\email{loock@uni-mainz.de}
\affiliation{Institute of Physics,
             Johannes Gutenberg University Mainz,
             Staudingerweg 7, 55128 Mainz, Germany}

\date{\today}

\begin{abstract}
In the framework of cavity QED, we propose a quantum repeater scheme that 
uses coherent light and atoms coupled to optical cavities. In contrast to 
conventional schemes, we exploit solely the cavity QED evolution for the 
entire quantum repeater scheme and, thus, avoid any explicit execution of 
quantum logical gates. The entanglement distribution between the repeater 
nodes is realized with the help of pulses of coherent light interacting 
with the atom-cavity system in each repeater node. In our previous paper 
[D.~Gonta and P.~van Loock, Phys. Rev. A \textbf{86}, 052312 (2012)], we 
already proposed a dynamical protocol to purify a bipartite entangled state 
using the evolution of atomic chains coupled to optical cavities. Here, we 
incorporate parts of this protocol in our repeater scheme, combining it 
with dynamical versions of entanglement distribution and swapping.
\end{abstract}


\maketitle

\section{Introduction}

In classical data transmission, repeaters are used to amplify the
data signals (bits) when they become weaker during their propagation
through the transmission channel. In contrast to classical
information, the above mechanism is impossible to realize when the
transmitted data signals are the carriers of quantum information
(qubits). In optical systems, for instance, a qubit is typically
encoded by means of a single photon which cannot be amplified or
cloned without destroying quantum information associated with this
qubit \cite{nat299, pla92}. Therefore, the photon has to propagate
along the entire length of the transmission channel which, due to
photon loss, leads to an exponentially decreasing probability to
receive this photon at the end of the channel.

To avoid exponential decay of a photon wave-packet and preserve
its quantum coherence, the quantum repeater was proposed \cite{prl81}.
This repeater contains three building blocks which have to be applied
sequentially. With the help of entanglement distribution, first, a large
set of low-fidelity entangled qubit pairs is generated between all repeater
nodes. Using entanglement purification, afterwards, high-fidelity
entangled pairs are distilled from this large set of low-fidelity entangled
pairs by means of local operations performed in each repeater node and
classical communication between the nodes \cite{prl76, prl77}.
Entanglement swapping, finally, combines two entangled pairs
distributed between the neighboring repeater nodes into one entangled
pair, thus, gradually increasing the distance of shared entanglement
\cite{swap}.

Because of the fragile nature of quantum correlations and inevitable
photon loss in an optical fiber, in practice, it poses a serious
challenge to outperform the direct transmission of photons along the
fiber. Up to now, however, only particular building blocks of an optical
quantum repeater have been experimentally demonstrated, for instance,
bipartite entanglement purification \cite{prl90, nat443}, entanglement
swapping \cite{prl96, pra71}, and entanglement distribution \cite{nat454,
sc316} between two neighboring nodes. Motivated both by the impressive
experimental progress and theoretical advances, various revised
and improved implementations of repeaters and their building blocks have
been recently proposed \cite{pra77, pra81, pra84a, rmp83, lpr}.

Practical and efficient schemes for implementing a quantum repeater
are not straightforward. The two mentioned protocols, entanglement
purification and entanglement swapping, in general, require 
feasible and reliable quantum logic, such as single- and two-qubit
logical gates. Because of the high complexity and demand of physical
resources, entanglement purification is the most challenging part of
a quantum repeater. The conventional purification protocols
\cite{prl77, pra59}, for instance, involve multiple applications of
controlled-NOT gates which assume sophisticated pulse sequences
posing thus a serious bottleneck for most physical realizations of
qubits \cite{nat443, prl104, pra71a, prl85, pra78, pra67}.

\begin{figure*}[!ht]
\begin{center}
\includegraphics[width=0.95\textwidth]{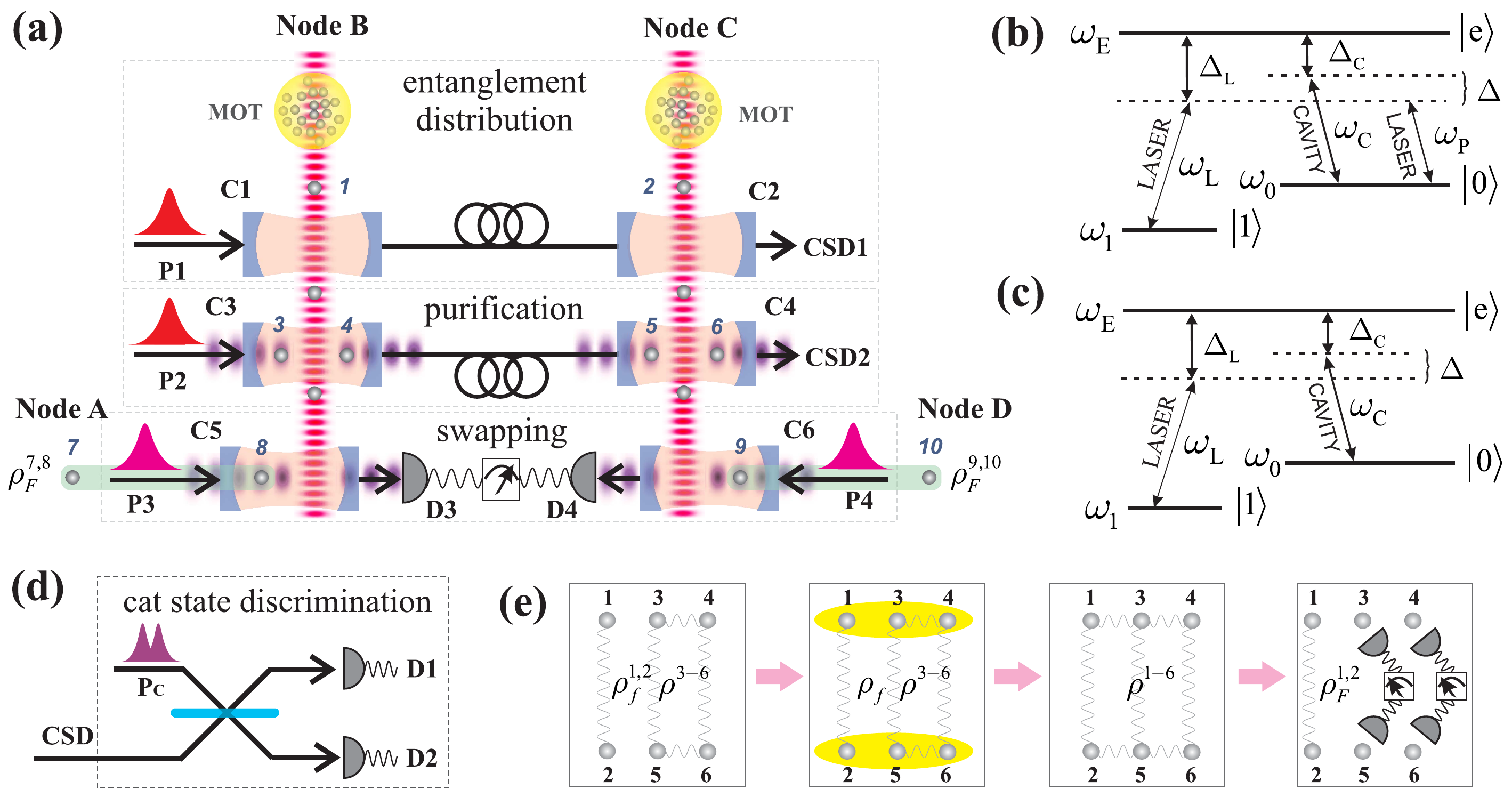} \\
\caption{(Color online) (a) Sketch of experimental setup that
realizes the two-node repeater scheme. (b, c) Structure of a
three-level atom in the $\Lambda$-configuration subjected to the
cavity and laser fields. (d) The cat state discrimination (CSD)
device. (e) Sequence of steps used in the purification protocol. See
text for description.}
\label{fig1}
\end{center}
\end{figure*}

In our previous papers \cite{pra84, pra86}, we already suggested a
practical scheme to purify dynamically a bipartite entangled state
by exploiting solely the evolution of short chains of atoms coupled
to high-finesse optical cavities. In the present paper, we make one
step further and propose an entire quantum repeater scheme that is
realized in the framework of cavity QED and incorporates
all three building blocks described above. In contrast to
conventional repeater schemes, we exploit solely the cavity QED
evolution and, thus, avoid completely quantum logical gates. The
entanglement distribution between the repeater nodes is realized
with the help of pulses of coherent light interacting sequentially
with the atom-cavity systems in each repeater node.

The paper is organized as follows. In the next section, we describe
in detail our dynamical quantum repeater scheme. We introduce and
discuss the entanglement distribution, purification, and swapping
protocols in Secs.~II.A, II.B, and II.C, respectively. In Sec.~II.D,
we discuss a few relevant issues related to the implementation of
our repeater scheme, while a brief rate analysis together with a 
summary and outlook are given in Sec.~III.

\section{Dynamical quantum repeater without quantum gates}

The main physical resources of our dynamical repeater are (i)
three-level atoms, (ii) high-finesse optical cavities, (iii)
continuous and pulsed laser beams, (iv) balanced beam splitters, and
(v) photon-number resolving detectors. In Fig.~\ref{fig1}(a) we
display the sketch of experimental setup that includes two repeater
nodes (B and C) and incorporates the entanglement distribution,
purification, and swapping protocols in one place.

In this setup, each repeater node includes single-mode cavities
$C_1$, $C_3$, and $C_5$ ($C_2$, $C_4$, and $C_6$), a chain of
equally distanced atoms conveyed along the setup with the help of an
vertical optical lattice, a pair of stationary atoms trapped inside
the cavity $C_3$ ($C_4$) by means of a horizontal optical lattice,
source of short coherent-state pulses $P_3$ ($P_4$), detector $D_3$
($D_4$) connected to the neighboring node via a classical
communication channel, and a magneto-optical trap (MOT) that plays
the role of source for the conveyed atoms. The alignment of vertical
lattice is such that the conveyed atoms cross cavities at their
anti-nodes ensuring, therefore, a strong atom-cavity coupling once
the atom is inside. In addition, the node B contains two
coherent-state pulse sources $P_1$ and $P_2$, while the node C
contains two cat state discrimination devices $CSD_1$ and $CSD_2$.
As shown in Fig.~\ref{fig1}(d), each such device includes the source
$P_C$ of single cat states (see below), a balanced beam splitter,
and two photon-number resolving detectors $D_1$ and $D_2$. Both
repeater nodes share two chains of atoms ($1$, $2$, and e.t.c.),
which are conveyed with a constant velocity through all three
cavities, and two pairs of stationary atoms $3$, $4$ and $5$, $6$
trapped inside the cavities $C_3$ and $C_4$, respectively. The atoms
$8$ and $9$ trapped inside the cavities $C_5$ and $C_6$,
respectively, are entangled to the atoms $7$ and $10$ trapped in the
repeater nodes A and D.

The setup in Fig.~\ref{fig1}(a) is divided in three (framed by dashed
rectangles) parts corresponding to the main building blocks of a quantum
repeater and mentioned in the introduction. Below, we relate step-by-step
our dynamical repeater scheme with this experimental setup and clarify
the role of each element.

\subsection{Entanglement Distribution}

The entanglement distribution protocol is shown in the top rectangle. In
this part of setup, the atoms are extracted one-by-one from the MOT,
initialized in the ground state $\ket{0}$, and inserted into the conveyor
such
that the atoms ($1$ and $2$) arrive the cavities $C_1$ and $C_2$ at the
same time. The state $\ket{0}$ together with the state $\ket{1}$ encode
a qubit by means of a three-level atom in the $\Lambda$-configuration as
displayed in Figs.~\ref{fig1}(b) and (c). In order to protect this qubit
against the decoherence, the states $\ket{0}$ and $\ket{1}$ are chosen as
the stable ground and long-living metastable atomic states or as the two
hyperfine levels of the ground state.

Once conveyed into the cavity $C_1$ ($C_2$), the atom $1$ ($2$)
couples simultaneously to the photon field of cavity and two
continuous laser beams as displayed in Fig.~\ref{fig1}(b). The laser
beams act vertically along each conveyor axis and are not depicted
in Fig.~\ref{fig1}(a) for simplicity. The evolution of coupled
atom-cavity-laser system in both repeater nodes is governed by the
Hamiltonian
\begin{equation}\label{ham-acl}
H_{ACL} = \frac{\hbar \, J_1}{2} \left( a + a^\dagger \right) \sigma^X ,
\end{equation}
where $\sigma^X$ is the respective Pauli operator in the basis
$\{\ket{0}, \ket{1} \}$ and $J_1$ is the atom-field coupling. We
show in Appendix~\ref{app1} that the above Hamiltonian is produced
deterministically in our setup assuming a strong driving of atom and
large atom-field detuning for both laser and cavity fields.
Moreover, this Hamiltonian implies that the (fast-decaying) excited
state $\ket{e}$ remains almost unpopulated during the evolution. The
evolution governed by $H_{ACL}$ yields the operator
\begin{equation}\label{evol-acl}
U_{ACL}(\alpha) = e^{\left( \alpha \, a^\dag - \alpha^* a \right) \, \sigma^X}
              = D \left( \alpha \, \sigma^X \right) \, ,
\end{equation}
where $\alpha = - \im J_1 \, t / 2$. This operation displaces the
cavity field mode by the amount $\alpha$ conditioned on the
atomic state. The complex amplitude $\alpha$ is
proportional to the atom-cavity-laser evolution time $t$ that, in
turn, is inverse proportional to the velocity of conveyed atom.

It has been suggested in Ref.~\cite{prl101} that the controlled
displacement provides an efficient scheme to distribute the
entanglement between two atoms coupled to remote cavities. We modify
this scheme by considering the feasible atom-cavity-laster evolution
(\ref{evol-acl}) controlled by $\sigma^X$ and using an input
(coherent-state) pulse $\beta$ generated by the source $P_1$ that
heralds the entanglement distribution. Our modified scheme works as
follows. First, the pulse $\beta$ interacts with the
atom-cavity-laser system in node B, where the cavity is prepared in
the vacuum state, while the atom is initialized in the ground state.
Assuming that $\beta^* = - \beta$, the evolution (\ref{evol-acl})
leads to the atom-pulse entangled state
\begin{equation}
U^B_{ACL} (\alpha) \ket{0_1} \ket{\beta}
      = \frac{1}{\sqrt{2}} \left( \ket{+_1} \ket{\beta + \alpha}
                                + \ket{-_1} \ket{\beta - \alpha} \right) \, . \notag
\end{equation}

Once the conveyed atom decouples from the cavity, the pulse in the 
state $\ket{\beta
\pm \alpha}$ leaks out of the cavity and is entirely outputted to
the transmission channel between the nodes. Since we are dealing
with a high-finesse cavity and since the fast-decaying atomic state
$\ket{e}$ remains almost unpopulated during the evolution, the
dominant photon loss occurs in the optical fiber that connects the
cavities $C_1$ and $C_2$ and that plays the role of transmission
channel in our setup. Apparently, the photon loss increases with the
length of the fiber. To a good approximation, therefore, this
observation suggests us to describe the loss using the beam splitter
model that transmits only a part of the pulse though the channel
\begin{equation}
\ket{\text{vac}}_E \, \ket{\beta \pm \alpha}_F \longrightarrow
 \ket{\sqrt{1 - \eta} \, (\beta \pm \alpha)}_E \, \ket{\tilde{\beta} \pm \tilde{\alpha}}_F \, ,
\end{equation}
where $\tilde{\beta} \pm \tilde{\alpha} = \sqrt{\eta} \, (\beta \pm
\alpha)$, while the subscripts $E$ and $F$ refer the environmental
and fiber light modes, respectively. Here $\eta = e^{-\ell /
\ell_\circ}$ describes the attenuation of the transmitted
(coherent-state) pulse through the fiber, where $\ell$ is the
distance between the repeater nodes, while $\ell_\circ$ is the
attenuation length that (for fused-silica fibers at
telecommunication wavelength) can reach almost $25$ km.

Next, the damped pulse interacts with the atom-cavity-laser system
in node C, where (as in the node B) the cavity is prepared in the
vacuum state, while the atom is initialized in the ground state. By
tracing over the environmental degrees of freedom (modes with the
subscript $E$), the evolution $U^B_{ACL} (\alpha)$ followed by
$U^C_{ACL} (\tilde{\alpha})$ leads to the mixed entangled state
between the both atoms and the coherent-state pulse
\begin{equation}\label{state0}
\rho = \frac{1 + e^{-2 |\alpha|^2 (1 - \eta)}}{2} \ket{p} \bra{p}
       + \frac{1 - e^{-2 |\alpha|^2 (1 - \eta)}}{2} \ket{h} \bra{h} \, ,
\end{equation}
where
\begin{eqnarray}\label{aaf}
\ket{p} &=& \frac{\sqrt{\mathcal{N}_+}}{2} \ket{\phi^+_{1,2}} \ketp{c}
            + \frac{\sqrt{\mathcal{N}_-}}{2} \ket{\psi^+_{1,2}} \ketm{c}
            + \frac{1}{\sqrt{2}} \ket{\phi^-_{1,2}} \ket{\tilde{\beta}} \, ; \notag \\
\ket{h} &=& \frac{\sqrt{\mathcal{N}_-}}{2} \ket{\phi^+_{1,2}} \ketm{c}
            + \frac{\sqrt{\mathcal{N}_+}}{2} \ket{\psi^+_{1,2}} \ketp{c}
            - \frac{1}{\sqrt{2}} \ket{\psi^-_{1,2}} \ket{\tilde{\beta}} \, ; \notag \\
&& \quad \ket{c_\pm} = \frac{1}{\sqrt{2 \, \mathcal{N}_\pm}}
                         \left( \ket{\tilde{\beta} + 2 \, \tilde{\alpha}} \pm
                                \ket{\tilde{\beta} - 2 \, \tilde{\alpha}} \right) \, ,
\end{eqnarray}
while $\mathcal{N}_\pm = 1 \pm e^{-8 \, \eta |\alpha|^2}$. The 
states $\ketp{c}$ and $\ketm{c}$ are the displaced (by the amount
$\tilde{\beta}$) even and odd Schr\"{o}dinger cat states,
respectively \cite{sch}.

The resulting (coherent-state) pulse leaks from $C_2$ and is
discriminated in the basis $\{ \ket{\tilde{\beta}}, \ketp{c},
\ketm{c} \}$ using the cat state discrimination device $CSD_1$
(see Sec.~II.D). Since $\ketm{c}$ is orthogonal to both $\ketp{c}$
and $\ket{\tilde{\beta}}$\footnote{To demonstrate this, one has to 
take into account the properties $\alpha^* = - \alpha$ and 
$\beta^* = - \beta$.}, we postselect only the detection events 
corresponding to this (odd) cat state. With the probability of 
success $\mathcal{N}_- / 4$, therefore, the entanglement 
distribution results into the rank 2 mixed state
\begin{equation}\label{state1}
\rho_f^{1,2} = f \, \ket{\psi^+_{1,2}} \bra{\psi^+_{1,2}}
               + (1-f) \ket{\phi^+_{1,2}} \bra{\phi^+_{1,2}} \, ,
\end{equation}
where the fidelity of entanglement is given by
\begin{equation}\label{fidelity}
f = \frac{1}{2} \left( 1 + \exp{\left[-2 |\alpha|^2 (1 - \eta) \right]} \right) \, .
\end{equation}
Once the output of measurements corresponds to the state
$ \ket{\tilde{\beta}}$ or $\ketp{c}$, the entanglement distribution is
unsuccessful. In this case, the atoms $1$ and $2$ should be discarded and
the entire sequence repeated using the next atomic pair conveyed from MOTs
in both repeater nodes. We remark, finally, that the fidelity (\ref{fidelity})
is close to the unity only when $|\alpha|^2 (1 - \eta) \ll 1$.
This observation suggests us to consider the values $|\alpha| \leq 1$ through
the paper.

\subsection{Entanglement Purification}

Assuming that the entanglement distribution is successful, the
(low-fidelity) entangled atoms $1$ and $2$ are conveyed along the
setup to the purification part displayed in the middle rectangle of
Fig.~\ref{fig1}(a). In this part of setup, each of cavities $C_3$
and $C_4$ share a pair of trapped atoms $3$, $4$ and $5$, $6$,
respectively. Both cavities are initially prepared in the vacuum
state, while the atoms are initialized in the state $\ket{0}$. The
entanglement purification is performed in three steps which are
displayed in Fig.~\ref{fig1}(e) and explained below.

\begin{figure}[!t]
\begin{center}
\includegraphics[width=0.45\textwidth]{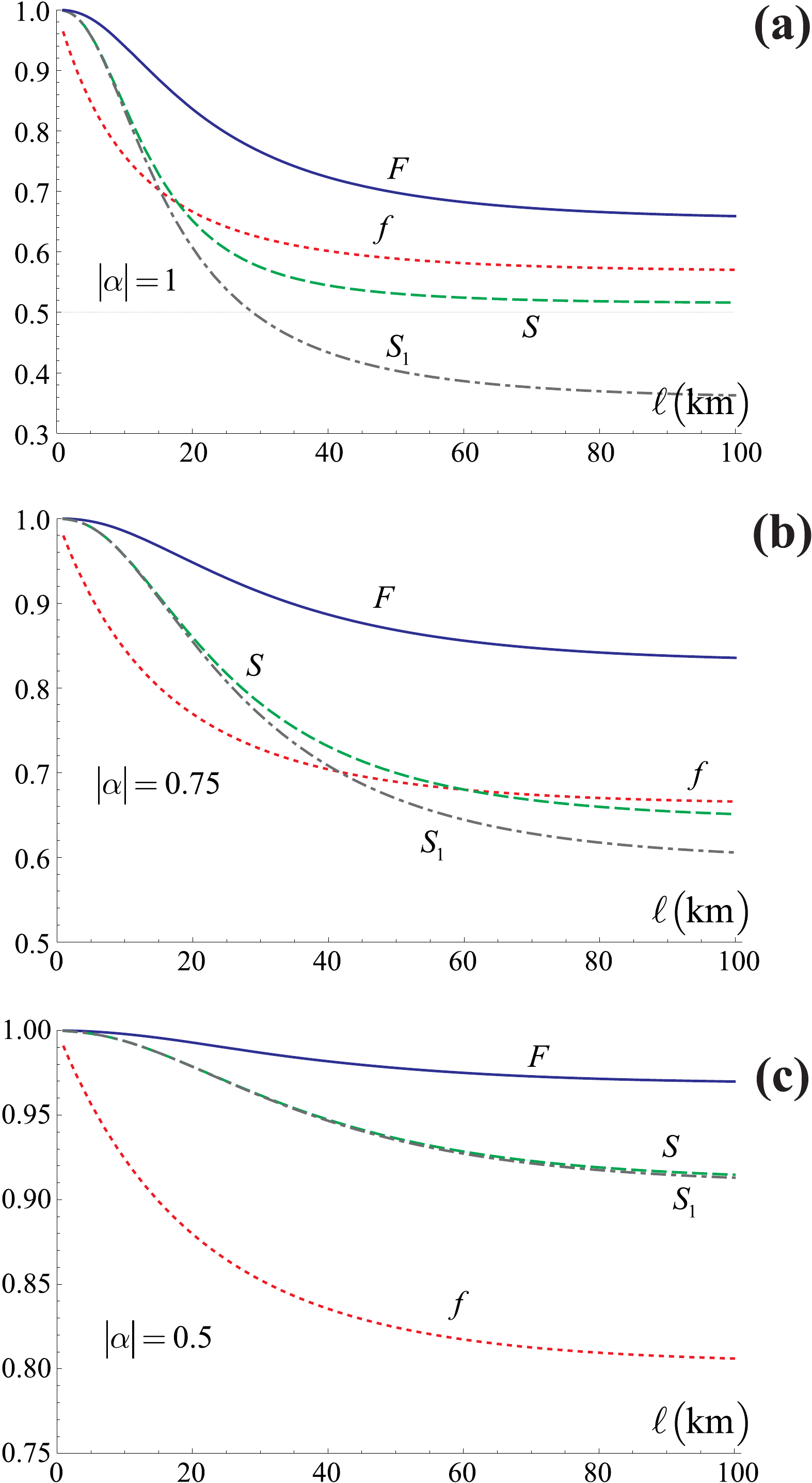} \\
\caption{(Color online) For the values (a) $|\alpha| = 1$, (b)
$|\alpha| = 0.75$, and (c) $|\alpha| = 0.5$, the solid and dotted
curves display the distribution and purification fidelities
(\ref{fidelity}) and (\ref{final1}), respectively. The dot-dashed
curves display the swapping fidelity $S_1$, while the dashed curves
show the fidelity (\ref{final2}) obtained in the conventional
swapping protocol.}
\label{fig2}
\end{center}
\end{figure}
\subsubsection{Four-qubit entanglement generation}

Right before atoms $1$ and $2$ are conveyed and coupled to the 
cavities $C_3$ and $C_4$, we generate the four-qubit entangled 
state
\begin{widetext}
\begin{eqnarray}\label{state2}
\rho^{3-6} &=& \frac{1}{4} \left( 1 - e^{- 8 |\alpha|^2 (1 - \eta )}  \right) 
               \ket{\phi^+_{3,4}, \phi^-_{5,6}} \bra{\phi^+_{3,4}, \phi^-_{5,6}} + 
               \frac{1}{4} \left( 1 + e^{- 8 |\alpha|^2 (1 - \eta )} \right) 
               \ket{\psi^+_{3,4}, \phi^-_{5,6}} \bra{\psi^+_{3,4}, \phi^-_{5,6}} \notag \\
           &+& \frac{1}{2} \, e^{- 8 |\alpha|^2 (1 - \eta )} 
               \ket{\phi^-_{3,4}, \psi^+_{5,6}} \bra{\psi^+_{3,4}, \phi^-_{5,6}} +
               \frac{1}{2} \, e^{- 8 |\alpha|^2 (1 - \eta )} 
               \ket{\psi^+_{3,4}, \phi^-_{5,6}} \bra{\phi^-_{3,4}, \psi^+_{5,6}} +
               \frac{1}{2} \, \ket{\phi^-_{3,4}, \psi^+_{5,6}} \bra{\phi^-_{3,4}, \psi^+_{5,6}} \, ,
\end{eqnarray}
\end{widetext}
associated with the atoms $3$, $4$, $5$, and $6$. This state is 
generated using the similar mechanism as utilized in the previous 
section. Namely, we employ sequentially the
evolutions $U_{2ACL}(\alpha)$ and $U_{2ACL}(\tilde{\alpha})$ in
repeater nodes B and C, respectively, where $(i = 3,4$ or $5,6)$
\begin{equation}\label{evol-acl2}
U_{2ACL}(\alpha) = e^{\left( \alpha \, a^\dag - \alpha^* a \right) \, \sum \sigma^X_i}
              = D \left( \alpha \, \sum \sigma^X_i \right) \, ,
\end{equation}
governed by the Hamiltonian
\begin{equation}\label{ham-acl2}
H_{2ACL} = \frac{\hbar \, J_1}{2} \left( a + a^\dagger \right) \sum \sigma^X_i \, .
\end{equation}
This Hamiltonian is produced deterministically in our setup (see
Appendix~\ref{app1}) and it describes the coupled system of atoms
$3$, $4$ ($5$, $6$), photon field of $C_3$ ($C_4$), and two
continuous laser beams which act vertically along each conveyor axis
(not depicted in Fig.~\ref{fig1}(a) for simplicity).

Similar to the previous section, the pulse $\beta$ (source $P_2$)
interacts sequentially with the atoms-cavity-laser systems in each
repeater node and is discriminated in the basis $\{
\ket{\tilde{\beta}}, \ketp{c}, \ketm{c}, \ketp{d}, \ketm{d} \}$ by
the $CSD_2$ (see Sec.~II.D), where

\begin{equation}
\ket{d_\pm} = \frac{1}{\sqrt{2 \left(1 \pm e^{-16 \, \eta |\alpha|^2} \right)}}
                         \left( \ket{\tilde{\beta} + 4 \, \tilde{\alpha}} \pm
                                \ket{\tilde{\beta} - 4 \, \tilde{\alpha}} \right) \, . \notag
\end{equation}
Only the events corresponding to the output state $\ketm{c}$ are
postselected which, with the probability of success $\mathcal{N}_- /
4$, result into the four-partite entangled state (\ref{state2}).

\subsubsection{Evolution for purification}

In the previous steps, we have successfully generated the entangled
states $\rho_f^{1,2}$ and $\rho^{3-6}$ [see the first rectangle of
Fig.~\ref{fig1}(e)]. In the next step, the atoms $1$ and $2$ are
conveyed and coupled to the cavities $C_3$ and $C_4$, respectively.
Since the coherent-state pulse ($P_2$) has already left the
cavities, they are both in the vacuum state. The conveyed atoms
together with the trapped atoms form two atomic triplets $1$, $3$,
$4$ and $2$, $5$, $6$. Each of these triplets evolves now due to the
periodic Heisenberg XY Hamiltonian \cite{ap16} $(i = 1,3,4$ or
$2,5,6)$
\begin{equation}\label{ham-xy}
H_{XY} = \frac{\hbar \, J_2}{2} \sum_i
         \left( \sigma_i^X \sigma_{i+1}^X + \sigma_i^Y \sigma_{i+1}^Y \right),
\end{equation}
over the time period $T_2 = 2 \, J_2^{-1}$,
where $\sigma^X_i$ and $\sigma^Y_i$ are the Pauli operators and $J_2$ is the
coupling between the atoms of a given triplet. We show in Appendix~\ref{app2}
that this Hamiltonian is produced deterministically in our scheme by coupling
simultaneously three atoms to the same cavity mode and a laser beam in the limit
of large detuning.

The evolution governed by the above Hamiltonian
\begin{equation}\label{evol1}
e^{- \frac{\im}{\hbar} H_{XY} \, t} = \sum_{j=1}^8 e^{- \frac{\im}{\hbar} E_j \, t}
                                      \, \ket{\mathbf{k}_j} \bra{\mathbf{k}_j} \, ,
\end{equation}
is completely determined by the energies $E_i$ and vectors
$\ket{\mathbf{k}_i}$, which satisfy the equality $H_{XY} \,
\ket{\mathbf{k}_i} = E_i \, \ket{\mathbf{k}_i}$ along with the
orthogonality $\braket{\mathbf{k}_i}{\mathbf{k}_j} = \delta_{ij}$
and completeness $\sum \ket{\mathbf{k}_i} \bra{\mathbf{k}_i} =
\vecI$ relations. With the help of Jordan-Wigner transformation
\cite{zp47}, this eigenvalue problem can be solved exactly (see, for
instance, Ref.~\cite{pra64}). Since the evolution operator
(\ref{evol1}) acts on the atomic triplet in node B that is entangled
with the atomic triplet in the node C, we consider the evolution
operator
\begin{equation}\label{evol2}
U_{XY} = \sum_{i,j =1}^8 e^{- \frac{\im}{\hbar} \left( E_i + E_j \right) \, T_2}
          \, \ket{\mathbf{k}^B_i} \bra{\mathbf{k}^B_i}
          \otimes \ket{\mathbf{k}^C_j} \bra{\mathbf{k}^C_j} \, ,
\end{equation}
referred to below as the purification and indicated by the
ellipses in Fig.~\ref{fig1}(e).

According to this evolution, the state of both atomic triplets
is described by the six-qubit density operator
\begin{eqnarray}\label{density5}
\rho^{1-6} &=& U_{XY}
               \left( \rho_{f}^{1,2} \otimes \rho^{3-6} \right) U^\dag_{XY} \notag \\
                            &=& \sum_{i,j=1}^{64} \, \rho^{1-6}_{i,j} \, \ketv{i} \brav{j} \, ,
\end{eqnarray}
where $2^6$ composite states $\ketv{i}$, satisfying the orthogonality and completeness
relations $\braketv{i}{j} = \delta_{ij}$ and $\sum \ketv{i} \brav{i} = \vecI$, respectively,
have been introduced. \footnote{Using the six-qubit density operator $\rho^{1-6}$,
we have routinely computed the matrix elements $\rho^{1-6}_{i,j}(T_2)$ from (\ref{density5})
which, however, are rather bulky to be displayed here.}

\subsubsection{Finalization of the protocol}

Once the purification is performed and the conveyed atoms leave
the respective cavities, the atoms $3$, $4$ and $5$, $6$ are
projected in the computational basis $\{ \ket{0}, \ket{1} \}$.
Entanglement purification is successful when the outcome of
projections coincides with one of combinations
\begin{subequations}\label{outcome}
\begin{eqnarray}
\{ 0_3, 1_4, 0_5, 1_6 \} ,  \quad \{ 0_3, 1_4, 1_5, 0_6 \} \, ; \\
\{ 1_3, 0_4, 0_5, 1_6 \} , \quad \{ 1_3, 0_4, 1_5, 0_6 \} \, .
\end{eqnarray}
\end{subequations}
With the constant probability of success $1/4$, therefore, the conveyed 
atoms are described by the density operator
\begin{eqnarray}\label{density6}
\rho^{1,2}_F &=& \frac{1}{\mathcal{N}_1} \sum_{\alpha, \beta = 1}^4
                       \rho^{1-6}_{\alpha, \beta} \, \ket{\mathbf{u}_\alpha} \bra{\mathbf{u}_\alpha} \notag \\
             &=& F \, \ket{\phi^-_{1,2}} \bra{\phi^-_{1,2}}
               + (1 - F) \ket{\psi^-_{1,2}} \bra{\psi^-_{1,2}} \, ,
\end{eqnarray}
where $\mathcal{N}_1$ is the respective normalization factor. In this 
expression, moreover, the purified fidelity is given by
\begin{widetext}
\begin{equation}\label{final1}
F = \frac{\left( 0.000548294 + 0.0016253 \, e^{\, 8 |\alpha|^2 (1 - \eta)} +
                          0.00217359  \, e^{\, 6 |\alpha|^2 (1 - \eta)} \right) f}
                          {0.000538502 + 0.00163509 \, e^{\, 8 |\alpha|^2 (1 - \eta)} +
                                                    \left( 0.00434718 \, f - 0.00217359 \right) e^{\, 6 |\alpha|^2 (1 - \eta)}} \, ,
\end{equation}
\end{widetext}
while $\ket{\mathbf{u}_\alpha}$ are determined by the one of 
inequalities
\begin{subequations}
\begin{eqnarray}
\ket{\mathbf{u}_\alpha} &\equiv& \langle 0_3, 1_4, 0_5, 1_6 | \mathbf{v}_\alpha \rangle \neq 0 \, ; \\
\ket{\mathbf{u}_\alpha} &\equiv& \langle 0_3, 1_4, 1_5, 0_6 | \mathbf{v}_\alpha \rangle \neq 0 \, ; \\
\ket{\mathbf{u}_\alpha} &\equiv& \langle 1_3, 0_4, 0_5, 1_6 | \mathbf{v}_\alpha \rangle \neq 0 \, ; \\
\ket{\mathbf{u}_\alpha} &\equiv& \langle 1_3, 0_4, 1_5, 0_6 | \mathbf{v}_\alpha \rangle \neq 0 \, ,
\end{eqnarray}
\end{subequations}
and correspond to the outcomes of the projections (\ref{outcome}).

The operator (\ref{density6}) implies that the entangled state
associated with the conveyed atoms $1$ and $2$ preserves its rank 2
form after the purification. Unlike the conventional purification
protocol, therefore, the purified state is completely characterized
by the fidelity (\ref{final1}). We stress that once the outcome of
atomic projections associated with the trapped atoms disagree with
(\ref{outcome}), the entanglement purification is unsuccessful. In
this case, the atoms $1$ and $2$ should be discarded and the entire
(repeater) sequence restarted using one fresh atomic pair conveyed
from MOTs in each repeater node.

By considering $|\alpha| = 1$, $0.75$, and $0.5$, in Fig.~\ref{fig2},
we compare the purification fidelity (solid curve) with the fidelity
(\ref{fidelity}) obtained by means of entanglement
distribution only (dotted curve). We see that purification
yields a significant growth of (input) distribution fidelity.
In agreement with Eqs.~(\ref{fidelity}) and
(\ref{final1}), moreover, these plots confirm that the smaller values
of $|\alpha|$ are chosen, the higher values of both distribution and
purification fidelities are obtained. In addition, we infer that
both fidelities saturate around $\ell \approx 100$ km and they exhibit
an almost constant behavior for $\ell > 100$ km.

We remark that the described purification protocol is based on the
effect of entanglement transfer between the networks of evolving
spin chains that was introduced and investigated in Ref.~\cite{qip}.
In the same reference, it was suggested that this effect plays the
key role in the entanglement distillation once a part of spins from
two such networks are projectively measured. One similar
entanglement purification protocol, moreover, has been proposed in
Ref.~\cite{pra78a}. In Refs.~\cite{pra84, pra86}, furthermore, we
adapted this mechanism to the cavity QED framework, where the role
of (spin-chain) networks was played by the atomic triplets coupled
to the cavities located in two repeater nodes, while the
cavity-mediated interaction (\ref{ham-xy}) reproduced the spin-chain
dynamics within a spin network. Using several improvements, in the
latter paper, we obtained an almost-unit output fidelity after a few
successful purification rounds. In contrast to Ref.~\cite{pra86},
however, in this paper we consider a notably modified approach to
the purification protocol and employ one single purification round.

\subsection{Entanglement Swapping}

Assuming that both the entanglement distribution and purification
were successful, the (high-fidelity) entangled atoms $1$ and $2$ are
conveyed along the setup to the swapping part displayed in the
bottom rectangle of Fig.~\ref{fig1}(a). In this part, the atoms $1$
and $2$ couple the cavities $C_5$ and $C_6$, both prepared initially
in the vacuum state. The conveyed atoms together with the trapped
atoms form two atomic pairs $1$, $8$, and $2$, $9$. We recall that
atoms $8$ and $9$ are entangled with atoms $7$ and $10$,
respectively, where each pair is described by the rank 2 mixed
states $\rho^{7,8}_F$ and $\rho^{9,10}_F$ given both by
(\ref{density6}).

\begin{table}
\caption{Bell states corresponding to $\mathbf{s}^1_{ij}$
for given $i$ and $j$.}
\begin{ruledtabular}
\begin{tabular}{c | c c c c}
\backslashbox{j}{i} & 1 & 2 & 3 & 4 \\ \hline
1 & $\phi^-_{7,10}$ & $\phi^+_{7,10}$ & $\psi^-_{7,10}$ & $\psi^+_{7,10}$ \\
2 & $\phi^+_{7,10}$ & $\phi^-_{7,10}$ & $\psi^+_{7,10}$ & $\psi^-_{7,10}$ \\
3 & $\psi^-_{7,10}$ & $\psi^+_{7,10}$ & $\phi^-_{7,10}$ & $\phi^+_{7,10}$ \\
4 & $\psi^+_{7,10}$ & $\psi^-_{7,10}$ & $\phi^+_{7,10}$ & $\phi^-_{7,10}$ \\
\end{tabular}
\end{ruledtabular}
\label{tab1}
\caption{Bell states corresponding to $\mathbf{n}^1_{ij}$
for given $i$ and $j$.}
\begin{ruledtabular}
\begin{tabular}{c | c c c c}
\backslashbox{j}{i} & 1 & 2 & 3 & 4 \\ \hline
1 & $\phi^+_{7,10}$ & $\phi^-_{7,10}$ & $\psi^-_{7,10}$ & $\psi^+_{7,10}$ \\
2 & $\phi^-_{7,10}$ & $\phi^+_{7,10}$ & $\psi^+_{7,10}$ & $\psi^-_{7,10}$ \\
3 & $\psi^+_{7,10}$ & $\psi^-_{7,10}$ & $\phi^-_{7,10}$ & $\phi^+_{7,10}$ \\
4 & $\psi^-_{7,10}$ & $\psi^+_{7,10}$ & $\phi^+_{7,10}$ & $\phi^-_{7,10}$ \\
\end{tabular}
\end{ruledtabular}
\label{tab2}
\end{table}

According to the conventional entanglement swapping protocol
\cite{swap}, the atomic pairs $1$, $8$ and $2$, $9$ are projectively
measured in the Bell bases $\ket{\mathbf{b}_{1,8}^i}$ and
$\ket{\mathbf{b}_{2,9}^i}$, respectively, where $(i = 1, \ldots, 4)$
\begin{equation}\label{basis1}
 \ket{\mathbf{b}_{a,b}^i}
   = \{ \ket{\phi^+_{a,b}}, \ket{\phi^-_{a,b}}, \ket{\psi^+_{a,b}}, \ket{\psi^-_{a,b}} \} \, .
\end{equation}
This projective measurement results unconditionally into the entangled
state between the (initially uncorrelated) atoms $7$ and $10$ increasing,
thus, the overall distance of shared entanglement from $\ell$ to $3 \, \ell$.
In other words, the swapped state is given by the density operator
\begin{eqnarray}\label{density7}
\rho^{7,10}_{S} &=& \frac{1}{\mathcal{N}_2} \bra{\mathbf{b}_{1,8}^i, \mathbf{b}_{2,9}^j}
                            \rho^{1,2}_{F} \otimes \rho^{7,8}_{F} \otimes \rho^{9,10}_{F}
                            \ket{\mathbf{b}_{1,8}^i, \mathbf{b}_{2,9}^j} \notag \\
                &=& S \, \ket{\mathbf{s}^1_{ij}} \bra{\mathbf{s}^1_{ij}} +
                      \left( 1-S \right) \ket{\mathbf{s}^2_{ij}} \bra{\mathbf{s}^2_{ij}} \, ,
\end{eqnarray}
where $\mathcal{N}_2$ is the respective normalization factor. 
In this expression, moreover, the resulting (swapped) fidelity 
takes the form
\begin{equation}\label{final2}
S = F (3 - 6 \, F + 4 \, F^2) \, ,
\end{equation}
while $\mathbf{s}^1_{ij}$ is displayed in Table~\ref{tab1}. Similar
to the states $\rho^{7,8}_F$, $\rho^{9,10}_F$, and $\rho^{1,2}_{F}$,
this swapped state $\rho^{7,10}_{S}$ preserves its rank 2 form and is
completely characterized by the fidelity (\ref{final2}) displayed
in Fig.~\ref{fig2} by dashed curves.

\begin{figure*}[!ht]
\begin{center}
\includegraphics[width=0.95\textwidth]{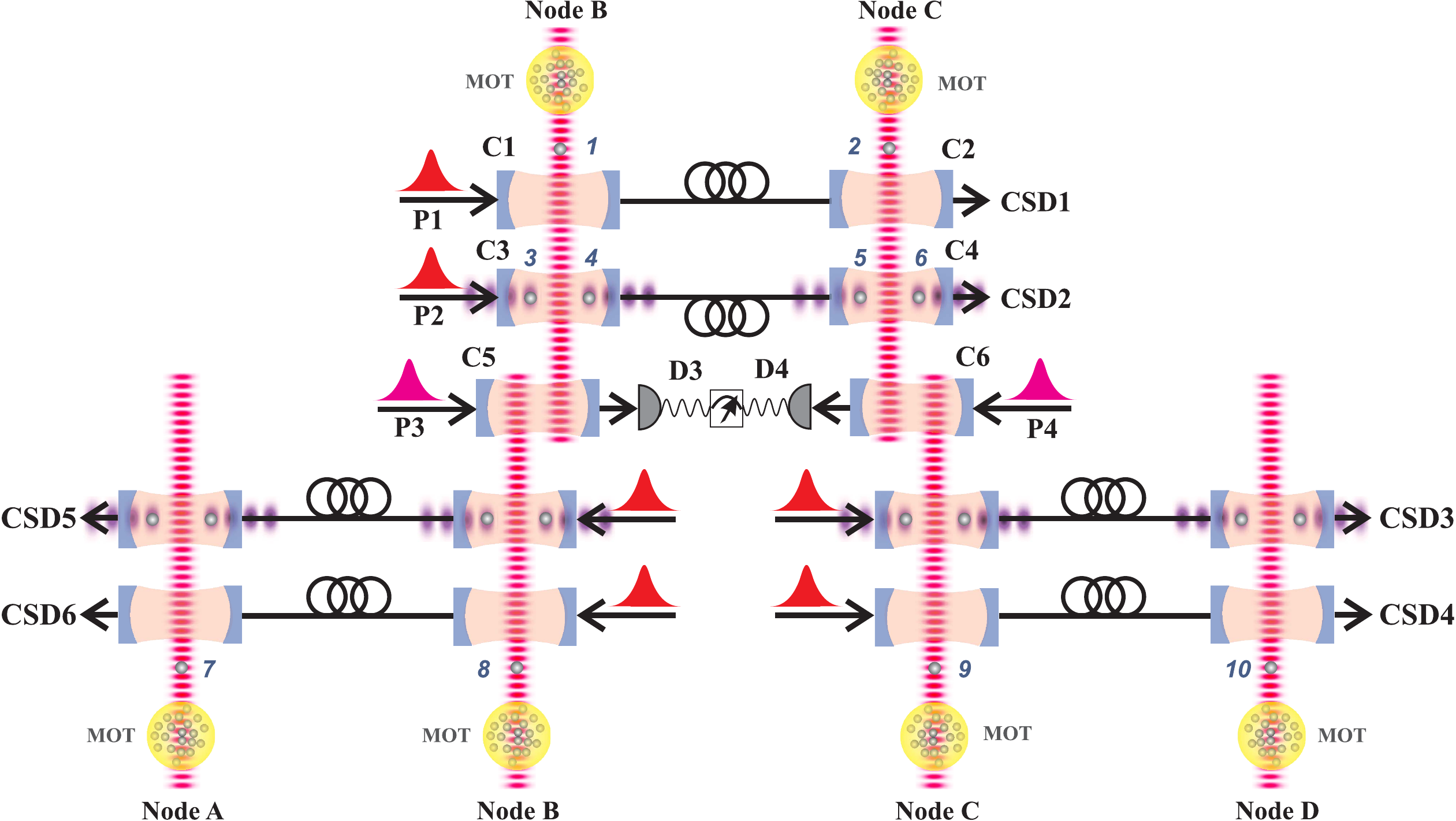} \\
\caption{(Color online) Sketch of experimental setup that realizes the
four-node repeater scheme.}
\label{fig3}
\end{center}
\end{figure*}

In our setup, each of atomic pairs evolves in cavity $C_5$ or $C_6$ due
to the periodic Heisenberg XX Hamiltonian $(i = 1,8$ or $2,9)$
\begin{equation}\label{ham-xx}
H_{XX} = \frac{\hbar \, J_2}{2} \sum_i \sigma_i^X \sigma_{i+1}^X \, ,
\end{equation}
over the time period $T_3 = \pi \, J_2^{-1} / 2$. We show in Appendix~\ref{app1}
that this Hamiltonian is produced deterministically in our setup by coupling
simultaneously two atoms to the same cavity mode and laser beams as displayed
in Fig.~\ref{fig1}(b). The evolution of an atomic pair governed by this Hamiltonian
over the period $T_3$ implies
\begin{subequations}\label{evol3}
\begin{eqnarray}
\ket{1_a, 1_b} &\longrightarrow& \frac{- \im}{\sqrt{2}} \left( \ket{0_a, 0_b} + \im \ket{1_a, 1_b} \right) \, ; \\
\ket{0_a, 0_b} &\longrightarrow& \frac{1}{\sqrt{2}} \left( \ket{0_a, 0_b} - \im \ket{1_a, 1_b} \right) \, ; \\
\ket{1_a, 0_b} &\longrightarrow& \frac{- \im}{\sqrt{2}} \left( \ket{0_a, 1_b} + \im \ket{1_a, 0_b} \right) \, ; \\
\ket{0_a, 1_b} &\longrightarrow& \frac{1}{\sqrt{2}} \left( \ket{0_a, 1_b} - \im \ket{1_a, 0_b} \right) \, , \quad
\end{eqnarray}
\end{subequations}
where the resulting states form the modified Bell basis
\begin{equation}
 \ket{\mathbf{m}_{a,b}^i} = e^{- \frac{\im}{\hbar} H_{XX} \, T_3}
              \{ \ket{1_a, 1_b}, \ket{0_a, 0_b}, \ket{1_a, 0_b}, \ket{0_a, 1_b} \} \, . \notag
\end{equation}

This evolution suggests an efficient and deterministic realization
of entanglement swapping in the framework of cavity QED. Namely, the
atomic pairs $1$, $8$ and $2$, $9$ are subjected to the evolution
$e^{- \frac{\im}{\hbar} H_{XX} \, T_3}$ followed by
projection in the computational basis $\{ \ket{0}, \ket{1} \}$.
Obviously, these two steps are equivalent with the projective
measurement in the modified Bell basis (\ref{evol3}), where the
swapped state is given by the expression
\begin{eqnarray}\label{density8}
\rho^{7,10}_{\mathbf{S}} &=& \frac{1}{\mathcal{N}_3}
            \bra{\mathbf{m}_{1,8}^i, \mathbf{m}_{2,9}^j}
            \rho^{1,2}_{F} \otimes \rho^{7,8}_{F} \otimes \rho^{9,10}_{F}
            \ket{\mathbf{m}_{1,8}^i, \mathbf{m}_{2,9}^j} \notag \\
              &=& S_1 \, \ket{\mathbf{n}^1_{ij}} \bra{\mathbf{n}^1_{ij}}
               + \sum_{k=2}^4 S_k \, \ket{\mathbf{n}^k_{ij}} \bra{\mathbf{n}^k_{ij}} \, ,
\end{eqnarray}
where $\mathcal{N}_3$ is the respective normalization factor, 
$\mathbf{n}^1_{ij}$ is displayed in Table~\ref{tab2}, while the 
functions
\begin{subequations}\label{final3}
\begin{eqnarray}
&& S_1 = F (1 - 2 \, F + 2 \, F^2) \, , \quad S_2 = 2 (1 - F) F^2 \, , \label{final4} \\
&& S_3 = 1 - 3 \, F + 4 \, F^2 - 2 \, F^3, \quad S_4 = 2 (F - 1)^2 F, \qquad
\end{eqnarray}
\end{subequations}
fulfill the inequality $S_1 \gg S_2 > S_3 > S_4$.
The swapped state $\rho^{7,10}_{\mathbf{S}}$ is diagonal in the standard
Bell basis (\ref{basis1}), where the function $S_1$ is identified with
the swapping fidelity $F_\text{final}$ and displayed in Fig.~\ref{fig2}
by dot-dashed curves. We see that for $|\alpha| > 0.5$ and $\ell < 20$ km,
this final fidelity almost coincides with the fidelity (\ref{final2})
obtained by means of conventional swapping protocol (see dashed curve).

The total probability of success associated with the entanglement
distribution, purification, and two swappings is given by the expression
\begin{equation}\label{prob}
P_\text{succ} = P^\text{sw}_\text{succ} \, \mathcal{N}^{\, 2}_- / 64
              = P^\text{sw}_\text{succ} \left( 1 - e^{-8 |\alpha|^2 \eta} \right)^2 / 64 \, ,
\end{equation}
where $P^\text{sw}_\text{succ}$ is the probability of success of
two entanglement swappings and is determined mainly by the detection
efficiency of atomic projective measurements (see below). For
$|\alpha| = 0.5$, 0.75, and 1, in Fig.~\ref{fig4}, we display the
ratio $P_\text{succ} / P^\text{sw}_\text{succ}$ as a function of
segment distance $\ell$. It seen that the total probability of
success is sensitive to the choice of $|\alpha|$ and it drops
dramatically with increasing $\ell$. This plot reveals the trade-off
between the length of a repeater segment and the total probability
of success.

\begin{figure}[!t]
\begin{center}
\includegraphics[width=0.45\textwidth]{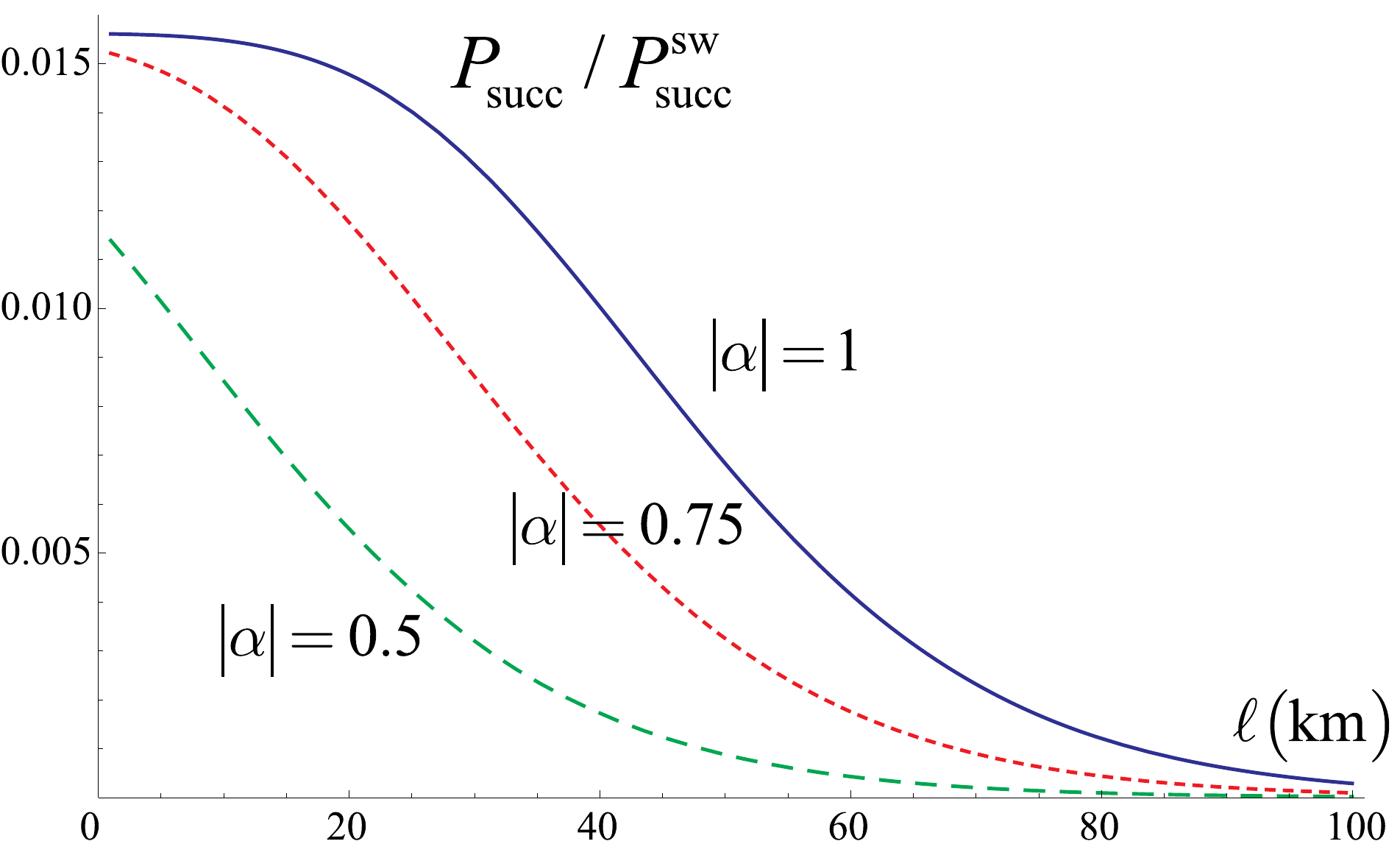} \\
\caption{(Color online) The ratio $P_\text{succ} / P^\text{sw}_\text{succ}$
as a function of repeater segment distance $\ell$ for $|\alpha| = 1$
(solid line), $|\alpha| = 0.75$ (dotted line), and $|\alpha| = 0.5$ (dashed
line).}
\label{fig4}
\end{center}
\end{figure}
\subsection{Remarks on the implementation of our scheme}

For simplicity, in the setup displayed in Fig.~\ref{fig1}(a), we
considered just two repeater nodes (B and C), where the atomic pairs
$7$, $8$ and $9$, $10$ have been initially entangled and given both
by (\ref{density6}). After we explained our repeater scheme, we are
ready to introduce the experimental setup that includes explicitly
nodes A, B, C, and D. This setup is displayed in Fig.~\ref{fig3}
and, in contrast to Fig.~\ref{fig1}(a), includes entanglement
distribution and purification protocols associated with the atomic
pairs $7$, $8$ and $9$, $10$, which are initially disentangled.
Simultaneously with the atomic pair $1$, $2$, these pairs are
conveyed along the setup (but in the opposite direction) and follow
the same sequence of cavity QED evolutions and atomic projective
measurements.

We recall that in the framework of entanglement distribution and
purifications protocols, the pulses $P_1$ and $P_2$ are
discriminated in the bases $\{ \ket{\tilde{\beta}}, \ketm{c},
\ketp{c} \}$ and $\{ \ket{\tilde{\beta}}, \ketm{c}, \ketp{c},
\ketm{d}, \ketp{d} \}$ once they leave the cavities $C_2$ and $C_4$,
respectively. In our setup, this discrimination is performed using
the cat state discrimination devices $CSD_1$ and $CSD_2$. As shown
in Fig.~\ref{fig1}(d), each such device includes a source $P_C$ of
single cat states $\ket{c_\pm}$, a balanced beam splitter, and two
photon-number resolving detectors $D_1$ and $D_2$.

We recall the property of cat states to interfere on a balanced beam 
splitter by producing an (amplitude) amplified cat state in one output 
mode and no photons in another output mode \cite{pra70}. This property 
suggest the following discrimination 
procedure. Once the leaked pulse leaves the respective cavity, it
interferes on the beam splitter with $\ket{c_+}$ or $\ket{c_-}$
generated by $P_C$. If both detectors $D_1$ and $D_2$ produce
clicks, then the leaked pulse is either $\ket{\tilde{\beta}}$ or
$\ket{d_\pm}$. However, if one of detectors produces no clicks, then
the state (leaving the other mode of beam splitter) is either an
even or odd (amplified) cat state. In contrast to an even cat state,
the odd cat state contains an odd number of photons on the top of
$|\sqrt{2} \, \tilde{\beta}|^2$. This feature plays a decisive role
in the discrimination between these cat states by means of the
photon-number resolving detector. We remark, finally, that the
generation of $\ket{c_\pm}$ by the source $P_C$ can be
deterministically realized using the cavity QED evolution $U_{ACL}(2
\, \tilde{\alpha})$ [see (\ref{evol-acl})] with an input pulse
$\ket{\tilde{\beta}}$, such that $\tilde{\beta}^* = -
\tilde{\beta}$. Assuming that the atom was initialized in the ground
state $\ket{0}$ and the cavity in the vacuum state, this evolution
results into the states $\ket{c_+}$ or $\ket{c_-}$ conditioned upon
the detection of atom in the state $\ket{0}$ or $\ket{1}$,
respectively.

We recall that all three building blocks of our repeater require an
efficient technique for projective measurements of atoms which are
trapped in (or conveyed through) a cavity. The method of atomic
non-destructive measurements demonstrated in Refs.~\cite{prl97,
prl103} enables projective measurements of single atoms coupled
(strongly) to a cavity field and fits perfectly in our experimental
setup. The physical mechanism behind these measurements exploits the
suppression of cavity transmission that arises due to the strong
atom-cavity coupling. Recall that each atom in our scheme is a
three-level atom in the $\Lambda$-configuration [see
Figs.~\ref{fig1}(b) and (c)], where only the states $\ket{0}$ and
$\ket{e}$ are coupled to the cavity field. If one such atom couples
the cavity and is prepared in the $\ket{0}$ state, such that the
cavity resonance is sufficiently detuned from the atomic $\ket{0}
\leftrightarrow \ket{e}$ transition frequency, then the cavity
transmission drops according to the atom-cavity detuning and
atom-cavity coupling. On the other hand, the cavity transmission
remains unaffected if the atom was prepared in the state $\ket{1}$.

Once sufficiently many readouts of the cavity transmission are
recorded, the mechanism described above enable us to determine the
state of a single atom with a reasonably high efficiency
\cite{prl97}. Since the atom-cavity coupling increases
proportionally with the number of loaded atoms, the same mechanism
enables us to distinguish the following three composite states of
two trapped atoms (i) $\ket{0_a, 0_b}$, (ii) $\ket{0_a, 1_b}$ or
$\ket{1_a, 0_b}$, and (iii) $\ket{1_a, 1_b}$ (see \cite{prl103}). We
remark, however, that this technique cannot distinguish between the
states $\ket{0_a, 1_b}$ and $\ket{1_a, 0_b}$ leading to an
incomplete knowledge about the swapped state (\ref{density8}) [see
Table~\ref{tab2}]. In order to avoid this drawback, one of the atoms
in $C_5$ ($C_6$) should decouple the cavity right after the
indecisive detection occurs and the entire measurement sequence
should be repeated with a single atom in the cavity. The decoupling
of an atom from the cavity, for instance, can be realized by
conveying one from the atoms further along the setup.

In addition, the projective measurements of atomic pairs $3$, $4$
and $5$, $6$ in cavities $C_3$ and $C_4$ are realized using a probe
beam (tuned to the cavity resonance frequency) produced by the
source $P_2$ and one of the photon detectors in $CSD_2$ that
monitors the transmission on the other end-point of the fiber.
During these projective measurements, the cat state source $P_C$ in
the upper input mode of beams splitter is switched off. In this
case, the probe beam includes the contributions from transmission of
both cavities. Recall that a successful purification event is
conditioned upon the combinations (\ref{outcome}) of atomic
projective measurements. These combinations imply that only one atom
in each cavity is excited. However, the combinations $\{ 0_3, 0_4,
1_5, 1_6 \}$ and $\{ 1_3, 1_4, 0_5, 0_6 \}$ lead to the same output
of signal transmission and have to be distinguished from the
combinations (\ref{outcome}). In order to discriminate this outcome
that implies a successful purification, the transmission
characteristics (i.e., the atom-cavity detuning and coupling)
associated with $C_3$ and $C_4$ should reasonably deviate.

\begin{figure}[!t]
\begin{center}
\includegraphics[width=0.45\textwidth]{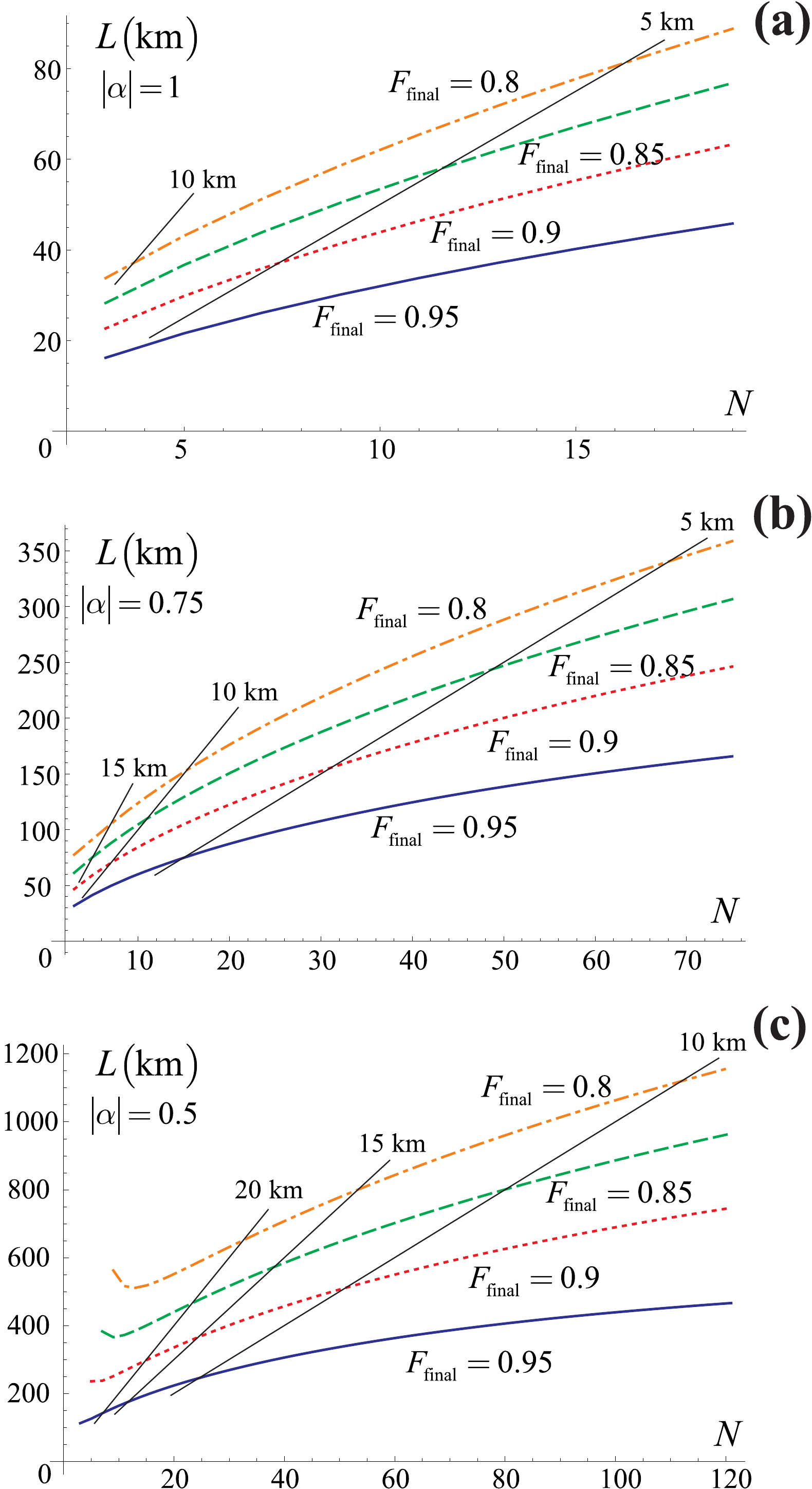} \\
\caption{(Color online) For the values (a) $|\alpha| = 1$, (b)
$|\alpha| = 0.75$, and (c) $|\alpha| = 0.5$, we display the overall
repeater length $L \equiv N \ell$ as a function of number of segments
$N$ plotted for the final fidelities $F_\text{final} = 0.95$ (thick
solid curve), $F_\text{final} = 0.9$ (dotted curve), $F_\text{final} = 0.85$
(dashed curve), and $F_\text{final} = 0.8$ (dot-dashed curve).
The thin solid lines reveal the maximally achievable length of
repeater for a given final fidelity corresponding to a constant
segment distance of $5$ km, $10$ km, $15$ km, or $20$ km. See text
for details.}
\label{fig5}
\end{center}
\end{figure}

Finally, the approach presented in this section requires that atoms
are transported with a constant velocity along the experimental
setup and coupled to the cavity and laser fields in a controllable
fashion. For this purpose, we introduced in our setup
magneto-optical traps (MOTs) which play the role of atomic source
and optical lattices (conveyors) which transport atoms with a
position and velocity control over the atomic motion. The proposed
setup is compatible with the existing experimental setups
\cite{prl95, prl98, njp10}, in which both MOTs and conveyors are
integrated into the same experimental framework with a high-finesse
optical cavity. The number-locked insertion technique \cite{njp12},
moreover, enables one to extract atoms from MOT and insert a
predefined pattern of them into an optical lattice with a
single-site precision. By encoding a qubit with the help of
hyperfine atomic levels, finally, it has been demonstrated that an
optical lattice preserves the coherence of this qubit over seconds
\cite{prl91, prl93}.

\section{Summary and Outlook}

\begin{table}
\caption{Various repeater characteristics for $|\alpha| = 1$.}
\begin{ruledtabular}
\begin{tabular}{c | c c c c}
$F_\text{final}$ & $\ell$ & $N$ & $R$ & $L$ \\ \hline
0.95 & 5.4 km & 3 & 39 pps & 16.3 km \\
0.9 & 5.1 km & 7  & 25 pps & 36 km \\
0.85 & 5.1 km & 11 & 19.2 pps & 56.4 km \\
0.8 & 5.2 km & 15 & 15.8 pps & 77.7 km \\
\end{tabular}
\end{ruledtabular}
\label{tab3}
\caption{Various repeater characteristics for $|\alpha| = 0.75$.}
\begin{ruledtabular}
\begin{tabular}{c | c c c c}
$F_\text{final}$ & $\ell$ & $N$ & $R$ & $L$ \\ \hline
0.95 & 5 km & 15 & 15.7 pps & 74.8 km \\
0.9 & 5 km & 31 & 10.2 pps & 155.2 km \\
0.85 & 5 km & 47 & 7.9 pps & 239 km \\
0.8 & 5 km & 67 & 6.5 pps & 337.6 km \\ \hline
0.95 & 10.6 km & 3 & 17.9 pps & 31.8 km \\
0.9 & 10 km & 7 & 11.6 pps & 70 km \\
0.85 & 10 km & 11 & 8.9 pps & 110 km \\
0.8 & 10.1 km & 15 & 7.4 pps & 151.8 km \\
\end{tabular}
\end{ruledtabular}
\label{tab4}
\caption{Various repeater characteristics for $|\alpha| = 0.5$.}
\begin{ruledtabular}
\begin{tabular}{c | c c c c}
$F_\text{final}$ & $\ell$ & $N$ & $R$ & $L$ \\ \hline
0.95 & 10 km & 25 & 3.4 pps & 247.5 km \\
0.9 & 10 km & 51 & 2.2 pps & 510.7 km \\
0.85 & 10 km & 79 & 1.7 pps & 796 km \\
0.8 & 10 km & 111 & 1.4 pps & 1115.3 km \\ \hline
0.95 & 15.5 km & 11 & 2.7 pps & 170.8 km \\
0.9 & 15.5 km & 23 & 1.8 pps & 356.4 km \\
0.85 & 15.3 km & 37 & 1.4 pps & 565 km \\
0.8 & 15 km & 53 & 1.2 pps & 798 km \\ \hline
0.95 & 20.2 km & 7 & 2.2 pps & 141.7 km \\
0.9 & 20 km & 15 & 1.5 pps & 298.4 km \\
0.85 & 20.2 km & 23 & 1.1 pps & 463.7 km \\
0.8 & 20.6 km & 31 & 0.9 pps & 639.1 km \\
\end{tabular}
\end{ruledtabular}
\label{tab5}
\end{table}

In the previous sections, we introduced our repeater scheme with
three segments (four nodes) corresponding to the overall distance $3
\, \ell$. The final fidelity $F_\text{final}$ was
identified with the swapping fidelity $S_1$ [see (\ref{final4})],
while the total probability of success was given by (\ref{prob}).
The probability of success $P^\text{sw}_\text{succ}$ associated with
this entanglement swapping is determined mainly by the detection
efficiency of atomic projective measurements \cite{prl97}.

The extension to an arbitrary amount of segments $N$ of the
(three-segment) repeater shown in Fig.~\ref{fig3} is straightforward.
For convenience, we consider odd values of $N$ corresponding to $N+1$
repeater nodes (or $N-1$ swapping operations). For
$|\alpha| = 1$, $0.75$, and $0.5$, we display in Fig.~\ref{fig5} the
dependence of the overall distance on the amount of repeater segments
taken for the final fidelities $F_\text{final} = 0.8$, $0.85$,
$0.9$, and $0.95$. The thin solid lines reveal the maximally achievable
overall distances along with the required amount of swappings
corresponding to the segment distances $\ell = 5$ km and $10$ km [in
Fig.~\ref{fig5}(a)] or $\ell = 10$ km, $15$ km, and $20$ km [in
Figs.~\ref{fig5}(b) and (c)]. As expected, in all three figures the segment
length $\ell$ decreases with the growing of $N$. This happens due to
the lack of (re)purification mechanism in our scheme that has to act
each time after the execution of several swappings.

Since small segment lengths lead to a rather dense distribution of repeater
nodes implying unreasonably high demand of physical resources and since
large segment lengths imply small probabilities of success (see
Fig.~\ref{fig4}), we intentionally bounded $N$ by the range $5$ km
$\leq \ell \leq 10$ km for $|\alpha| = 1$ and $10$ km $\leq \ell \leq 20$
km for $|\alpha| = 0.75$ and $|\alpha| = 0.5$. We observe, furthermore,
that small values of $|\alpha|$ along with small values of final fidelities
$F_\text{final}$ lead both to high overall distances $L \equiv N \ell$ of
repeater. For $|\alpha| = 0.5$ and $\ell = 10$ km, for instance, our repeater
distributes one entangled pair over the distance of almost $1115$ km with
$F_\text{final} = 0.8$ and $P_\text{succ} / P^\text{sw}_\text{succ} \approx
0.0085$. For $|\alpha| = 1$, the same segment length, and the same final
fidelity, in contrast, the entanglement is distributed over the distance of
almost $34$ km with $P_\text{succ} / P^\text{sw}_\text{succ} \approx 0.015$.

Besides the analysis of the final fidelities and probabilities of success,
we compute the repeater rates which provide together the main characteristics
of a quantum repeater. Since the atomic (fast-decaying) excited state remains
unpopulated and the qubit is encoded by means of long-living atomic states,
we assume that the coherence of atoms and cavities exceeds the overall time
required to complete the entanglement distribution, purification, and swapping
protocols in all repeater nodes. This assumption corresponds to
a repeater with ideal memory and implies that the main source of decoherence
is the photon loss in the optical fiber \cite{pra83}. Since the
probability of success is rather small in our scheme (see Fig.~\ref{fig4}),
we compute the repeater rates (in units of pairs per second) using the 
expression \cite{rmp83}
\begin{equation}\label{rate}
R = \left( \frac{2}{3} \right)^n \frac{P_\text{succ}}{T_\circ} \, ,
\end{equation}
where $T_\circ$ is the time required to distribute and purify an entangled
state over a single repeater segment being followed by two swappings, $P_\text{succ}$
is the probability of success (\ref{prob}), while $n$ is given by the equality
$2^{\, n} = N$.

Since the cavity based atomic measurements operate with a high efficiency
\cite{prl97}, to a good approximation, we set $P^\text{sw}_\text{succ} = 0.9$.
In accordance with the experimental setup shown in Fig.~\ref{fig1}(a), moreover,
we set $T_\circ = 7 \, \ell / \tilde{c}$, where $\tilde{c} = 2 \cdot 10^8$ m$/$s
is the speed of light in the optical fiber. The rate (\ref{rate}) is determined
by the triplet $\{ |\alpha|, \ell, N \}$ that we extract from Fig.~\ref{fig5} for
a given value of the final fidelity $F_\text{final}$. For instance, for $L = 33$
km, $|\alpha| = 1$, and $F_\text{final} = 0.8$, we find that $\ell = 11$ km, 
$N = 3$, and $P_\text{succ} = 0.014$. Being inserted in Eq.~(\ref{rate}), these 
values imply $R \approx 18$ pairs per second (pps). In a similar fashion, we 
display in Tables~\ref{tab3}, \ref{tab4}, and \ref{tab5} the rates calculated 
for various values of $\ell$, $F_\text{final}$, and $|\alpha|$.

In this paper, a fully cavity QED-based quantum repeater including 
entanglement distribution, purification, and swapping protocols was
proposed. In contrast to conventional repeater schemes, we completely 
avoid the explicit use of quantum logical gates by exploiting solely
cavity QED evolution. Our repeater scheme has a conveyor structure
design, in which a chain of initialized single atoms is inserted
into an optical lattice and conveyed along the entire repeater node.
At the same time, another chain of initialized atoms is conveyed
along the neighboring repeater node in a synchronous fashion. These
two nodes form together a repeater segment, while the entire set of
segments form the quantum repeater itself. Each atomic chain is
conveyed through the entanglement distribution and purification
blocks, such that each synchronized atomic pair becomes entangled
and (afterwards) purified in a probabilistic fashion. Finally, the
purified atomic pair is conveyed into the entanglement swapping
block, where two entangled atomic pairs distributed between the
neighboring repeater nodes are deterministically combined into one
entangled pair distributed over a longer distance.

A detailed experimental setup was proposed in Figs.~\ref{fig1}(a),
\ref{fig3} and a complete description of all necessary steps
and manipulations was provided. A comprehensive analysis of the
final fidelities obtained after multiple swapping operations was
performed and the correlation between the overall and the segment
distances was determined by means of Fig.~\ref{fig5}. Moreover, a
rate analysis has been performed and the main repeater characteristics
have been revealed in Tables~\ref{tab3}, \ref{tab4}, and \ref{tab5}.
Following recent developments in cavity QED, moreover, we briefly
pointed to and discussed a few practical issues related to the 
implementation of our purification scheme, including the main 
limitation that arises due to the lack of (re)purification mechanism. 
We stress that although the proposed quantum repeater is experimentally 
feasible, its explicit realization for a long-distance quantum 
communication still poses a serious challenge.

Finally, we like to mention Ref.~\cite{prl101} by Munro and co-authors
in which an entanglement distribution scheme was proposed. In contrast 
to our approach based on the displacement operator (\ref{evol-acl}) and 
controlled by $\sigma^X$, the approach of Munro and 
co-authors is based on the displacement controlled by $\sigma^Z$ which 
appears less feasible in the framework of cavity QED. A practical 
consequence of using the $\sigma^X$ displacement instead of $\sigma^Z$ 
is that the remote atoms have to be initialized 
in the ground state and not in an equal superposition of both basis states 
as in the approach of Munro and co-authors. In our scheme, moreover, we 
used an input (coherent-state) pulse $\beta$ generated by the source 
$P_1$ that heralds the generation of state (\ref{state0}).

\begin{acknowledgments}

We thank the BMBF for support through the QuOReP program.

\end{acknowledgments}

\appendix

\section{Derivation of the Hamiltonians \\
         (\ref{ham-acl}), (\ref{ham-acl2}), and (\ref{ham-xx})}\label{app1}

In this appendix, we show that the Hamiltonians (\ref{ham-acl}),
(\ref{ham-acl2}), and (\ref{ham-xx}) are produced deterministically
in our setup. Specifically, $N$ (three-level) atoms are subjected
simultaneously to the field of (initially empty) cavity and fields
of two laser beams as displayed in Fig.~\ref{fig1}(b). The evolution
of this coupled atoms-cavity-laser system is governed by the
Hamiltonian ($k=1,2, \ldots, N$)
\begin{eqnarray}\label{ham1}
H_1 &=& \hbar \, \omega_C \, a^\dag \, a - \im \hbar \sum_{k}
       \left[ \frac{g}{2} \, a \, \ket{e}_k \bra{0} \right. \\
       &+& \left. \frac{\Omega}{2} \left(
       e^{-i \omega_L \, t} \ket{e}_k \bra{1} +
       e^{-i \omega_P \, t} \ket{e}_k \bra{0}\right) - H.c. \right] \notag \\
       &+& \hbar \sum_{k} \left[
       \omega_1 \ket{1}_k \bra{1} +
       \omega_E \ket{e}_k \bra{e} +
       \omega_0 \ket{0}_k \bra{0} \right] \, , \notag
\end{eqnarray}
where $g$ denotes the coupling strength of atoms to the cavity mode,
while $\Omega$ denotes the coupling strengths of atoms to both laser
fields.

We switch to the interaction picture using the unitary transformation
\begin{equation}\label{picture1}
U_1 =    e^{- \im t \left[\sum \left(
         \omega_1 \ket{1}_k \bra{1} + \left( \widetilde{\omega} + \Delta_L \right)
         \ket{e}_k \bra{e} + \omega_0 \ket{0}_k \bra{0} \right)
         + \left( \widetilde{\omega} - \omega_0 \right) \, a^\dag a \right]}. \notag
\end{equation}
where $\widetilde{\omega} \equiv \omega_1 + \omega_L$.
In this picture, the Hamiltonian (\ref{ham1}) takes the form
\begin{eqnarray}\label{ham2}
H_2 &=& \hbar \, \Delta \, a^\dag a - \im \hbar \sum_{k}
       \left[ \frac{g}{2} \, e^{-i \Delta_L \, t} a \, \ket{e}_k \bra{0} \right. \notag \\
    &+& \left. \frac{\Omega}{2} \, e^{-i \Delta_L \, t} \left(
       \ket{e}_k \bra{1} + \ket{e}_k \bra{0}\right) - H.c. \right] .
\end{eqnarray}
where the notation $\Delta_L \equiv (\omega_E - \omega_1) - \omega_L$,
$\Delta_C \equiv (\omega_E - \omega_0) - \omega_C$, and
$\Delta \equiv \Delta_L - \Delta_C$ has been introduced.

We require that $\Delta_L$ and $\Delta_C$ are sufficiently far detuned,
such that no atomic $\ket{e} \leftrightarrow \ket{0}$ or $\ket{e} \leftrightarrow \ket{1}$
transitions can occur. We expand the evolution governed by the
Hamiltonian (\ref{ham2}) in series and keep the terms up to the second order,
\begin{equation}
U_2 \cong \vecI - \frac{\im}{\hbar} \int_{0}^{t} H_2 \, dt^\prime -
                    \frac{1}{\hbar^2} \int_{0}^{t} \left( H_2 \, \int_{0}^{t^\prime}
                    H_2 \, dt^{\prime \prime} \right) dt^\prime \, . \notag
\end{equation}
By performing integration and retaining only linear-in-time contributions,
we express this evolution in the form
\begin{equation}\label{operator2}
U_2 \cong \vecI - \frac{\im}{\hbar} \, H_3 \, t
    \cong \exp \left[ - \frac{\im}{\hbar} \, H_3 \, t \right],
\end{equation}
where the effective Hamiltonian is given by
\begin{equation}
H_3 = \hbar \, \Delta \, a^\dag a + \frac{\hbar \, \Omega}{4 \Delta_L} \sum_{k} \left[
      \Omega \, \ket{1}_k \bra{0} + g \, \ket{1}_k \bra{0} \, a + H.c. \right] \notag
\end{equation}
after removing constant contributions. We switch to
the interaction picture with respect to the first term of $H_3$.
In this picture, we obtain
\begin{equation}\label{ham4}
H_4 = \frac{\hbar \, \Omega}{4 \Delta_L} \sum_{k} \left[
      \Omega \, \ket{1}_k \bra{0} + g \, e^{-i \Delta \, t} \ket{1}_k \bra{0} \, a + H.c. \right] \, .
\end{equation}

We switch now from the atomic basis $\{ \ket{0}, \ket{1} \}$
to the basis $\{ \ket{+}, \ket{-} \}$, where
\begin{equation}\label{basis}
\ket{+} = \frac{1}{\sqrt{2}} \left( \ket{0} + \ket{1} \right);  \quad
\ket{-} = \frac{1}{\sqrt{2}} \left( \ket{0} - \ket{1} \right).
\end{equation}
In this basis, the Hamiltonian (\ref{ham4}) takes the form
\begin{eqnarray}\label{ham5}
H_5 &=& \frac{\hbar \, \Omega}{8 \Delta_L} \sum_{k} \left[ 2 \, \Omega S^Z_k
         + g \left( S^Z_k ( e^{-i \Delta \, t} a + e^{i \Delta \, t} a^\dag ) \right. \right. \notag \\
    && \hspace{1.5cm} \left. \left. + (S^\dag_k - S_k)(e^{-i \Delta \, t} a - e^{i \Delta \, t} a^\dag) \right)
     \right] , \qquad
\end{eqnarray}
where $S_k \equiv \ket{-}_k \bra{+}$ and $S^Z_k \equiv \ket{+}_k \bra{+} - \ket{-}_k \bra{-}$,
and where we removed all the constant contributions. We switch again to the
interaction picture with respect to the first term of (\ref{ham5}). In this
picture, we obtain
\begin{eqnarray}\label{ham6}
H_6 &=& \hbar \, \frac{g \, \Omega}{8 \Delta_L} \sum_{k}
        \left[ S^Z_k ( e^{-i \Delta \, t} a + e^{i \Delta \, t} a^\dag ) \right. \\
    && \left. + (S^\dag_k \, e^{\im \frac{\Omega^2}{2 \Delta_L} t}
    - S_k \, e^{-\im \frac{\Omega^2}{2 \Delta_L} t})(e^{-i \Delta \, t} a - e^{i \Delta \, t} a^\dag) \right] . \notag
\end{eqnarray}

In the strong driving regime, i.e., for $\Omega^2 / (2\, \Delta_L) \gg \{ \Delta, \,
g \, \Omega / (8 \Delta_L) \}$, we eliminate
the last (fast oscillating) term using the same arguments as for the rotating wave
approximation. Using the identity $S^Z_k = \sigma^X_k$, the Hamiltonian (\ref{ham6})
reduces to
\begin{equation}\label{ham7}
H_7 = \hbar \, \frac{g \, \Omega}{8 \Delta_L}
      \left( e^{-i \Delta \, t} a + e^{i \Delta \, t} a^\dag \right) \sum_{k} \sigma^X_k \, .
\end{equation}
In the case of vanishing $\Delta$ (equivalently $\Delta_L = \Delta_C$), the above
Hamiltonian takes the simplified form
\begin{equation}\label{ham8}
H_8 = \hbar \, \frac{g \, \Omega}{8 \Delta_L} \left(a + a^\dag \right) \sum_{k} \sigma^X_k \, ,
\end{equation}
which, under the notation $J_1 \equiv g \, \Omega / (4 \, \Delta_L)$, coincides with
the Hamiltonian (\ref{ham-acl}) for $N=1$ and with the Hamiltonian (\ref{ham-acl2})
for $N=2$.

In the case $\Delta_L \neq \Delta_C$, furthermore, we require that $\Delta$ is
sufficiently far detuned and expand the evolution governed by (\ref{ham7})
in series up to the second order. By performing integration and retaining only
linear-in-time contributions, we express this evolution in the form (\ref{operator2}),
where the resulting (effective) Hamiltonian
\begin{equation}
H_9 = \frac{\hbar \, g^2 \Omega^2}{64 \, \Delta_L^2 \, \Delta}
      \sum_{i,j}^{i \neq j} \left[ \sigma_i^\dag \sigma_j^\dag + \sigma_i^\dag \sigma_j
                             + \sigma_i \, \sigma_j^\dag + \sigma_i \, \sigma_j \right]
\end{equation}
coincides with the Hamiltonian (\ref{ham-xx}) under the notation
$J_2 \equiv g^2 \Omega^2 / (16 \, \Delta_L^2 \, \Delta)$, where
$\sigma_i \equiv \ket{0}_k \bra{1}$.

\section{Derivation of the Hamiltonian (\ref{ham-xy})}\label{app2}

In this appendix, we show that the Hamiltonian (\ref{ham-xy}) is
produced deterministically in our setup. Specifically, three
(three-level) atoms are subjected simultaneously to the field of
(initially empty) cavity and field of a laser beam as displayed in
Fig.~\ref{fig1}(c). The evolution of this coupled atoms-cavity-laser
system is governed by the Hamiltonian ($k=1,2,3$)
\begin{eqnarray}\label{ham10}
H_{10} &=& \hbar \, \omega_C \, a^\dag \, a \\
       &-& \im \hbar \sum_{k}
       \left[ \frac{g}{2} \, a \, \ket{e}_k \bra{0} + \frac{\Omega}{2} \,
       e^{-i \omega_L \, t} \ket{e}_k \bra{1} - H.c. \right] \notag \\
       &+& \hbar \sum_{k} \left[
       \omega_1 \ket{1}_k \bra{1} +
       \omega_E \ket{e}_k \bra{e} +
       \omega_0 \ket{0}_k \bra{0} \right] \, , \notag
\end{eqnarray}
where $g$ denotes the coupling strength of atoms to the cavity mode,
while $\Omega$ denotes the coupling strength of atoms to the laser
field.

We switch to the interaction picture using the unitary transformation
\begin{equation}
U_3 =    e^{- \im t \left[\sum \left(
         \omega_1 \ket{1}_k \bra{1} + \omega_E \ket{e}_k \bra{e}
         + \omega_0 \ket{0}_k \bra{0} \right)
         + \left(\omega_1 + \omega_L - \omega_0 \right) a^\dag a \right]}. \notag
\end{equation}
In this picture, the Hamiltonian (\ref{ham10}) takes the form
\begin{eqnarray}\label{ham11}
H_{11} &=& \hbar \, \Delta \, a^\dag \, a \\
       &-& \im \hbar \sum_{k}
       \left[ \frac{g}{2} \, a \, e^{i \Delta_L \, t} \ket{e}_k \bra{0} + \frac{\Omega}{2} \,
       e^{i \Delta_L \, t} \ket{e}_k \bra{1} - H.c. \right], \notag
\end{eqnarray}
where the notation $\Delta_L \equiv (\omega_E - \omega_1) - \omega_L$,
$\Delta_C \equiv (\omega_E - \omega_0) - \omega_C$, and
$\Delta \equiv \Delta_L - \Delta_C$ has been introduced.

We require that $\Delta_L$ and $\Delta_C$ are sufficiently far detuned, such that no atomic
$\ket{e} \leftrightarrow \ket{0}$ and $\ket{e} \leftrightarrow \ket{1}$ transitions can occur.
We expand the evolution governed by the Hamiltonian (\ref{ham11}) in series up to the second
order. By performing integration and retaining only linear-in-time contributions, we express
this evolution in the form (\ref{operator2}), where the effective Hamiltonian is given by (we
assume that the cavity field is initially in the vacuum state)
\begin{equation}\label{ham12}
H_{12} = \hbar \, \Delta \, a^\dag \, a
      + \hbar \, \frac{g \, \Omega}{4 \, \Delta_L} \sum_{k} \left[ a \, \ket{1}_k \bra{0} + H.c. \right] \, .
\end{equation}
We switch one more time to the interaction picture
with respect to the first term of (\ref{ham12}). In this interaction picture,
the resulting Hamiltonian takes the form
\begin{equation}\label{ham13}
H_{13} = \hbar \, \frac{g \, \Omega}{4 \, \Delta_L}
      \sum_{k} \left[ a \, e^{-i \Delta \, t} \ket{1}_k \bra{0} + H.c. \right] \, .
\end{equation}

We require, finally, that $\Delta$ is sufficiently far detuned. As above, we expand again
the evolution governed by the Hamiltonian (\ref{ham13}) in series up to the second order and
retain only linear-in-time contributions after the integration. This leads to the effective
Hamiltonian
\begin{equation}\label{ham14}
H_{14} = \frac{\hbar \, g^2 \Omega^2}{16 \, \Delta_L^2 \, \Delta}
      \left[\sum_{i,j}^{i \neq j} \ket{0_i, 1_j} \bra{1_i, 0_j} + \sum_k \ket{1}_k \bra{1} \right] \, .
\end{equation}
Since the second term in this Hamiltonian commutes with the first term, we
eliminate the second term by means of an appropriate interaction picture. The resulting
Hamiltonian, i.e., the first term of (\ref{ham14}), coincides with the Hamiltonian
(\ref{ham-xy}) under the notation $J_2 \equiv g^2 \Omega^2 / (16 \, \Delta_L^2 \, \Delta)$.


\begin{thebibliography}{50}

%
%
%

\bibitem{nat299}
 W.~K.~Wootters and W.~H.~Zurek,
 Nature \textbf{299}, 802 (1982).

\bibitem{pla92}
 D.~Dieks,
 Phys. Lett. A \textbf{92A}, 271 (1982).

\bibitem{prl81}
 H.-J.~Briegel, W.~D\"{u}r, J.~I.~Cirac, and P.~Zoller,
 Phys. Rev. Lett. \textbf{81}, 5932 (1998).

\bibitem{prl76}
 C.~H.~Bennett, G.~Brassard, S.~Popescu, B.~Schumacher, J.~A.~Smolin, and W.~K.~Wootters,
 Phys. Rev. Lett. \textbf{76}, 722 (1996).

\bibitem{prl77}
 D.~Deutsch, A.~Ekert, R.~Jozsa, C.~Macchiavello, S.~Popescu, and A.~Sanpera,
 Phys. Rev. Lett. \textbf{77}, 2818 (1996).

\bibitem{swap}
 M.~Zukowski, A.~Zeilinger, M.~A.~Horne, and A.~K.~Ekert,
 Phys. Rev. Lett. \textbf{71}, 4287 (1993).

%
%
%

\bibitem{prl90}
 Z.~Zhao, T.~Yang, Y.-A.~Chen, A.-N.~Zhang, and J.-W.~Pan,
 Phys. Rev. Lett. \textbf{90}, 207901 (2003).

\bibitem{nat443}
 R.~Reich et al.,
 Nature \textbf{443}, 838 (2006).

\bibitem{prl96}
 T.~Yang et al.,
 Phys. Rev. Lett. \textbf{96}, 110501 (2006).

\bibitem{pra71}
 H.~de~Riedmatten et al.,
 Phys. Rev. A \textbf{71}, 050302(R) (2005).

\bibitem{nat454}
 Z.-S.~Yuan, Y.-A.~Chen, B.~Zhao, S.~Chen, J. Schmiedmayer, and J.-W.~Pan,
 Nature \textbf{454}, 1098 (2008).

\bibitem{sc316}
 C.-W.~Chou et al.,
 Science \textbf{316}, 1316 (2007).

%
%
%

\bibitem{pra77}
 Y.-B. Sheng, F.-G. Deng, and H.-Y. Zhou,
 Phys. Rev. A \textbf{77}, 042308 (2008).

\bibitem{pra81}
 B.~Zhao, M.~M\"{u}ller, K.~Hammerer, and P.~Zoller,
 Phys. Rev. A \textbf{81}, 052329 (2010).

\bibitem{pra84a}
 C.~Wang, Y.~Zhang, and G.-S.~Jin,
 Phys. Rev. A \textbf{84}, 032307 (2011).

\bibitem{rmp83}
 N.~Sangouard, C.~Simon, H.~de~Riedmatten, and N.~Gisin,
 Rev. Mod. Phys. \textbf{83}, 33 (2011).

\bibitem{lpr}
 P.~van~Loock,
 Laser Photonics Rev. \textbf{5}, 167 (2011).

%
%
%

\bibitem{pra59}
 W.~D\"{u}r, H.-J.~Briegel, J.~I.~Cirac, and P.~Zoller,
 Phys. Rev. A \textbf{59}, 169 (1999).

%
%
%

\bibitem{prl104}
 L.~Isenhower et al.,
 Phys. Rev. Lett. \textbf{104}, 010503 (2010).

\bibitem{pra71a}
 A.~G\'{a}bris and G.~S.~Agarwal,
 Phys. Rev. A \textbf{71}, 052316 (2005).

\bibitem{prl85}
 S.-B.~Zheng and G.-C.~Guo,
 Phys. Rev. Lett. \textbf{85}, 2392 (2000).

\bibitem{pra78}
 T.~Tanamoto, K.~Maruyama, Y.-X.~Liu, X.~Hu, and F.~Nori,
 Phys. Rev. A \textbf{78}, 062313 (2008).

\bibitem{pra67}
 N.~Schuch and J.~Siewert,
 Phys. Rev. A \textbf{67}, 032301 (2003).

%
%
%

\bibitem{pra84}
 D.~Gonta and P.~van~Loock,
 Phys. Rev. A \textbf{84}, 042303 (2011).

\bibitem{pra86}
 D.~Gonta and P.~van~Loock,
 Phys. Rev. A \textbf{86}, 052312 (2012).

%
%
%

\bibitem{prl101}
 W.~J.~Munro, R.~Van~Meter, S.~G.~R.~Louis, and K.~Nemoto,
 Phys. Rev. Lett. \textbf{101}, 040502 (2008).

%
%
%

\bibitem{sch}
 E.~Schr\"{o}dinger,
 Naturwissenschaften \textbf{23}, 807 (1935).

\bibitem{ap16}
 E.~Lieb, T.~Schultz, and D.~Mattis,
 Ann. Phys. (N.Y.) \textbf{16}, 407 (1961).

%
%
%

\bibitem{zp47}
P.~Jordan and E.~Wigner,
Z. Phys. \textbf{47}, 631 (1928).

\bibitem{pra64}
 X.~Wang,
 Phys. Rev. A \textbf{64}, 012313 (2001).

%
%
%

\bibitem{qip}
 A.~Casaccino, S.~Mancini, and S.~Severini,
 Quant. Inf. Process. \textbf{10}, 107 (2011).

\bibitem{pra78a}
 K.~Maruyama and F.~Nori,
 Phys. Rev. A \textbf{78}, 022312 (2008).

\bibitem{pra70}
 A.~P.~Lund, H.~Jeong, T.~C.~Ralph, and M.~S.~Kim,
 Phys. Rev. A \textbf{70}, 020101(R) (2004).

%
%
%

\bibitem{prl97}
 A.~D.~Boozer, A.~Boca, R.~Miller, T.~E.~Northup, and H.~J.~Kimble,
 Phys. Rev. Lett. \textbf{97}, 083602 (2006).

\bibitem{prl103}
 M.~Khudaverdyan, W.~Alt, T.~Kampschulte, S.~Reick, A.~Thobe, A.~Widera, D.~Meschede,
 Phys. Rev. Lett. \textbf{103}, 123006 (2009).

%
%
%

\bibitem{prl95}
 S.~Nussmann et al.,
 Phys. Rev. Lett. \textbf{95}, 173602 (2005).

\bibitem{prl98}
 K.~M.~Fortier, S.~Y.~Kim, M.~J.~Gibbons, P.~Ahmadi,
 and M.~S.~Chapman,
 Phys. Rev. Lett. \textbf{98}, 233601 (2007).

\bibitem{njp10}
 M.~Khudaverdyan et al.,
 New J. Phys. \textbf{10}, 073023 (2008).

%
%
%

\bibitem{njp12}
 M.~Karski et al.,
 New J. of Phys. \textbf{12} 065027 (2010).

\bibitem{prl91}
 S.~Kuhr et al.,
 Phys. Rev. Lett. \textbf{91}, 213002 (2003).

\bibitem{prl93}
 D.~Schrader, I.~Dotsenko, M.~Khudaverdyan, Y.~Miroshnychenko,
 A.~Rauschenbeutel, and D.~Meschede,
 Phys. Rev. Lett. \textbf{93}, 150501 (2004).

%
%
%

\bibitem{pra83}
 N.K.~Bernardes, L.~Praxmeyer, and P.~van~Loock,
 Phys. Rev. A \textbf{83}, 012323 (2011).

%
%
%

\end{thebibliography}
\end{document}